# Five Statistical Questions about the Tree of Life

David J. Aldous[1,*], Maxim A. Krikun[2], and Lea Popovic[3]

[1]*Department of Statistics, University of California, Berkeley, 367 Evans Hall # 3860, Berkeley, CA 94720-3860, USA;* [2]*Institut Elie Cartan, Universite Henri Poincare, Nancy, France; and* [3]*Department of Mathematics and Statistics, Concordia University, Montreal, Canada H3G 1M8*
*Correspondence to be sent to: Department of Statistics, University of California, Berkeley,
367 Evans Hall # 3860, Berkeley, CA 94720-3860, USA; E-mail: aldous@stat.berkeley.edu.*

*Abstract.*—Stochastic modeling of phylogenies raises five questions that have received varying levels of attention from quantitatively inclined biologists. 1) How large do we expect (from the model) the ratio of maximum historical diversity to current diversity to be? 2) From a correct phylogeny of the extant species of a clade, what can we deduce about past speciation and extinction rates? 3) What proportion of extant species are in fact descendants of still-extant ancestral species, and how does this compare with predictions of models? 4) When one moves from trees on species to trees on sets of species (whether traditional higher order taxa or clades within PhyloCode), does one expect trees to become more unbalanced as a purely logical consequence of tree structure, without signifying any real biological phenomenon? 5) How do we expect that fluctuation rates for counts of higher order taxa should compare with fluctuation rates for number of species? We present a mathematician's view based on an oversimplified modeling framework in which all these questions can be studied coherently. [Diversification; macroevolution; neutral model; phylogeny; speciation; stochastic.]

There is a long tradition, dating back at least to Yule (1924), of study of simple stochastic models for aspects of *biodiversity* or *macroevolution*—the evolutionary history of speciations and extinctions. Such models have several distinct potential uses.

- As null models when seeking to attribute biological significance to data. Is a particular clade expansion an instance of adaptive radiation (Bond and Opell 1998)? Does a typical clade expansion represent expansion into novel niches or replacement of other clades (Benton 1999)? Were speciation rates unusually high during the Cambrian radiation (Lieberman 2001)? Are the apparent "pulses" of extinctions and speciations seen in the fossil record of large mammals in southern and eastern Africa over the last 3 myr real, or could they be an artifact arising from the limited number of different dates of sites yielding the fossils (McKee 1995, 2001)?
- Detecting analogs of the regression fallacy in which some observed effect (predictable on statistical grounds) is incorrectly thought to require some causal explanation.
- Among several approaches to reconstructing phylogenies on extant species from molecular data, the Bayesian approach requires a prior distribution on possible trees (Huelsenbeck et al. 2002).

In this paper, we present five questions that we hope will interest quantitatively inclined biologists and describe the predictions made by certain probability models. The first three questions are studied using the Aldous and Popovic (2005) model for a clade with $n$ extant species, described in the next section, and the final two questions use the induced models for phylogenetic trees on genera introduced in Aldous et al. (2008), described later. We quote analytic results from those papers and give new simulation results but no new analytic results.

Over the last 15 years, a huge range of stochastic models for specific aspects of evolution related to our questions have been studied, and we make no attempt to mention them all. Nee (2004, 2006) provides a useful overview of the relevant biology literature.

As an alternative to potentially more realistic but incompatible models for different aspects, in a "back to basics" approach we use a simple neutral model of species-level macroevolution plus models of how biologists might assign species to higher level taxa. This is all set up so that questions about both tree shape and time series of taxa, at both species level and higher levels, for both extant and extinct clades, can be addressed *within the same framework*. In contrast, no previous modeling framework we have seen is capable of addressing all five questions we pose.

An ultimate goal of probability modeling is to identify which features of observed data might have arisen by chance and which require biological explanation. But this is valuable only if one believes that the chance models are realistic, in the sense of plausible null hypotheses. The authors of this paper are mathematicians, and our intention is to present models for the consideration of biologists, not to address the more substantial issues of realism of models. We focus on a neutral model because there the mathematics is simple enough to permit analytic results. As will be discussed in the conclusion, one could replace our species-level model with some other model while retaining our methodology for extension to higher order taxa.

## Overview of Model

A standard model (Nee 2006) is the linear birth-and-death model in which each species becomes extinct at rate (probability per unit time) μ and has daughter species at rate λ. Our focus is on the case where λ = μ. In mathematical terminology, this is the *critical* case; biologically one could call it the *neutral* case, meaning there is no assumption that diversity intrinsically tends to increase. The *supercritical* case λ > μ, leading to exponential growth in diversity, has quite different mathematical behavior, discussed briefly in the conclusion. Choosing the time unit to be mean species lifetime, the critical case has λ = μ = 1. In other words, species lifetime has the exponential (mean 1) distribution, and during the lifetime, daughter species arise after independent exponential (mean 1) random times.

The following *conditioned critical branching process* model, discussed in Aldous and Popovic (2005) and further discussed in Gernhard (2008), for the phylogenetic tree on a clade with *n* extant species (for given *n*) is intended to capture the intuitive idea of "purely random macroevolution."

The Bayesian terminology (prior, posterior) is convenient but keep in mind we are just making a mathematical definition, not doing statistical analysis.

1. The clade originates with one species at a random time before present, whose prior distribution is uniform on $(0, \infty)$.
2. As time runs forward, diversity evolves as the linear birth-and-death model with λ = μ = 1.
3. Condition on the number of species at the present time $t = 0$ being exactly equal to *n*.

The "posterior distribution" on the evolution of lineages given this conditioning now yields a probability distribution on phylogenetic trees on *n* extant species and a variable number of extinct species. Note that this is a "complete" tree in which lineages of extant species pass through explicit extinct species, but of course removing this information gives what we will call the "lineage tree" on the *n* extant species (and this is what is usually called the "phylogenetic tree" on extant species). Figure 1 shows some realizations of the model.

Our first three questions will be studied using this species-level model, but the final two involve higher order taxa (for concreteness of language we write *genera*) and will be studied using schemes introduced in Aldous et al. (2008) for extending the underlying species-level model to trees whose terminal taxa are clades. This extension, explained later, involves a numerical parameter θ (the chance a new species is a "new type") and one of three *classification schemes* for partitioning species into genera so that (to a different extent in each of the schemes) new types are founders of new genera.

## Past Fluctuations in Sizes of Extant Clades

**Question 1.** Within a typical clade, how do we expect current diversity to compare with the maximum historical diversity?

Readers will be familiar with the following intuitively appealing biological explanation of readily identifiable clades. A successful clade begins with a *key innovation* in one species, followed by a rapid *adaptive radiation* of species sharing that innovation; clade size increases until a level set by ecological constraints and stays at roughly this maximum level (while individual species arise and disappear) until some extrinsic factor upsets the equilibrium. Notwithstanding textbook examples of clades (horse, rhinoceros) that were much larger in the past, some version of this "logistic" picture is often taken to be self-evident, as the following quote indicates:

> [We study] theoretical clades that have either been growing exponentially throughout their history or have been of constant size, such that each time a new lineage has appeared by speciation another lineage has gone extinct. *These extremes bracket the plausible dynamical histories of real clades.* . . . Logistic growth, in which diversity rises to some maximum, is a convenient model for macroevolutionary clade expansion . . . . In this framework, exponential growth is the early phase of logistic growth, and the constant size model describes a clade that has been at its maximum size for some time. (Nee and May 1997, p. 692; our emphasis added)

This may be a perfectly reasonable view of large clades (flowering plants, birds, mammals), but what about small clades? Let us quantify Question 1 by considering the statistic

$$R = \frac{\text{maximum number of species at any one past time}}{\text{current number of species}}.$$

Here $R \geq 1$ because we include "current time" in "any past time." How large should we expect $R$ to be? Our species-level model (Aldous and Popovic 2005, Corollary 6) predicts a $1/r$ law

$$\Pr(R \geq r) = 1/r, \quad 1 \leq r < \infty,$$

so that $R$ would vary widely between clades, with a median value of 2. This contrasts with the view, implicit in the quotation above, that typically $R$ will be close to 1. It would be interesting to attempt to estimate the distribution of $R$ from the fossil record.

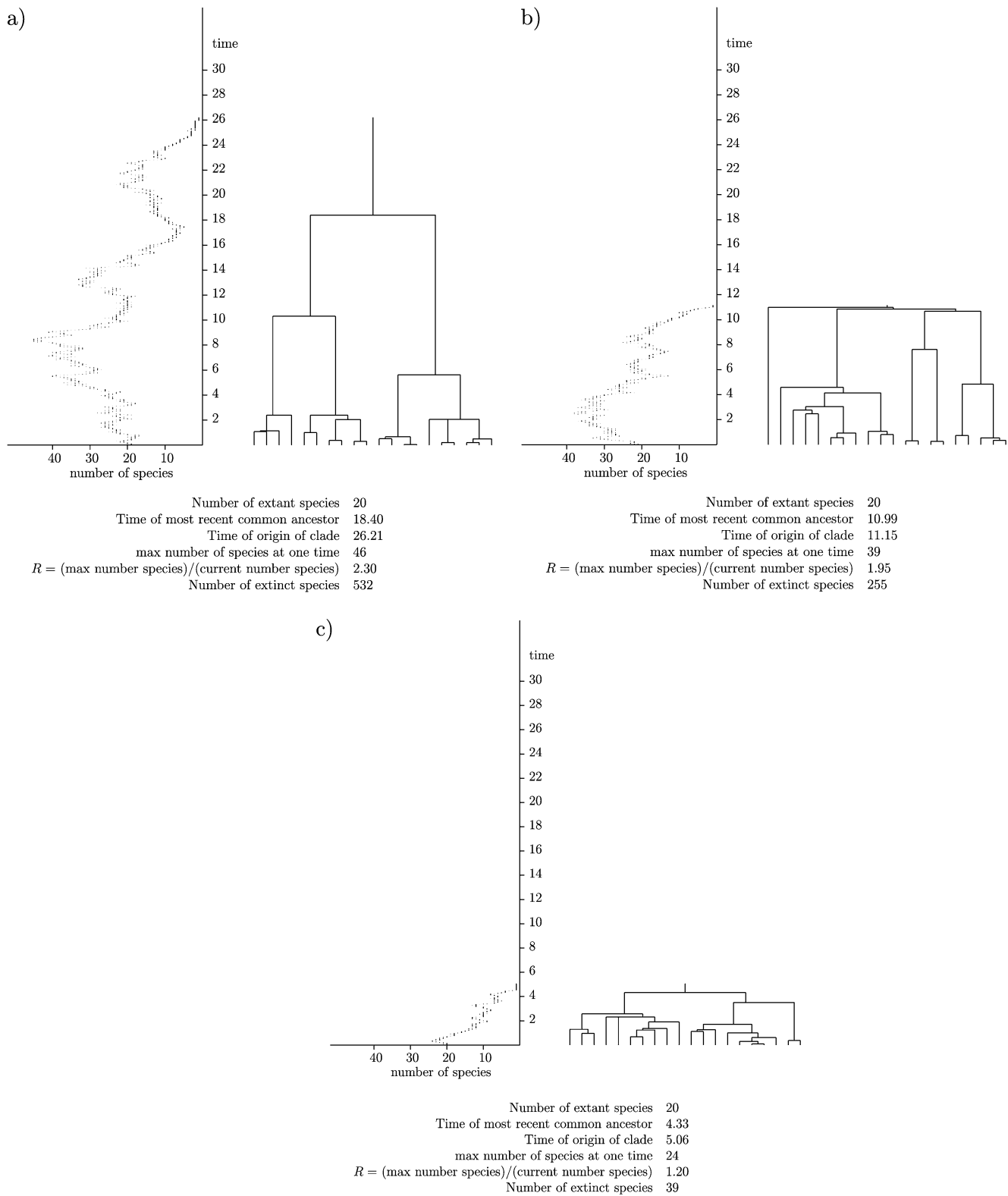

FIGURE 1. Three tree realizations (a, b, and c) of the lineage tree in our model, $n = 20$.

### Variability of Realizations and Conservative Analysis of Biological Significance

Though the intrinsic variability of realizations of stochastic models of macroevolution has often been noted, the classical statistical practice of comparing *averages* of quantities derived from data with averages predicted by models often makes it hard to keep variability in mind. Let us illustrate one aspect of variability. Consider three quantities associated with our species model: the time $T_n^{\text{origin}}$ of clade origin; the time $T_n^{\text{mrca}}$ of the most recent common ancestor of extant species; and the number $Z_n$ of species at the time of most recent common ancestor (recall the time unit equals mean species lifetime). It turns out that these quantities grow linearly with $n$, but there is no "law of averages"; instead as $n \to \infty$, there is a limit joint distribution, say $(T^{\text{origin}}, T^{\text{mrca}}, Z)$, for the normalized statistics $(n^{-1}T_n^{\text{origin}}, n^{-1}T_n^{\text{mrca}}, n^{-1}Z_n)$, and an explicit formula for this limit joint distribution can be found in Aldous and Popovic (2005, Corollary 8).

Figure 2 illustrates this by showing 10 realizations from the *limit* joint distribution. The key point to note is that each of these quantities varies by a factor of 10 over the realizations.

A complete tree from our species process is easy to simulate using the structure of the time-reversed process given in Aldous and Popovic (2005, Lemma 8). Three realizations of the lineage tree on $n$ extant species for $n = 20$ are shown in Figure 1, demonstrating vividly that different realizations can look very different. If we saw three real trees with such radically different radiation patterns and times, then we would surely be inclined to attribute biological significance to the differences. This recognition of variability is central to any perspective on the following question.

**Question 2.** From a correct phylogeny of the extant species of a clade, what can we deduce about past speciation and extinction rates?

This question is already much discussed in the biological literature; we will just make one qualitative comment and one quantitative comment.

### Qualitative Assessments of Nonrandomness

Wollenberg et al. (1996) show three published examples of lineage trees (columbines, cranes, the *Drosophila virilis* group: our comments refer to the B examples) that look quite different. After doing a statistical test of significance, with reference to a probability model that is similar to ours but with realizations with large fluctuations of species count censored, they conclude (p. 842)

> ... consistent with the original authors' impressions, the columbines and the cranes do indeed display nonrandom phylogenetic patterns of diversification, with the columbines showing recent and the cranes showing ancient significant clustering of speciation events ... [also consistent were] the results for the *D. virilis* group, where no evidence of temporal nonrandomness [could be identified].

But the visual difference between these three empirical trees is no larger than the visual difference between realizations of our model—indeed, the three realizations in Figure 1 resemble the (recent, ancient, steady) diversification visible for (columbines, cranes, *D. virilis*)—and we suspect that any reasonable test statistic would indicate that each data tree from Wollenberg et al. (1996) might plausibly arise "by chance" within our model. The point is that our model, with its intrinsic greater variability, provides a *more conservative* approach to assessing significance of observed features of phylogenetic trees.

### Estimating Rates in Birth-and-Death Processes: A Simulation Study

More concretely, consider the problem of estimating past speciation and extinction rates within a clade using only the lineage tree (assumed correct) on extant species. An often used model is the linear birth-and-death model, which we now regard as having three parameters $(t_*, \lambda, \mu)$, where

$t_* = $ time before present of clade origin,

$\lambda i = $ total speciation rate, when $i$ species,

$\mu i = $ total extinction rate, when $i$ species.

It is routine to calculate numerically maximum likelihood estimates (MLEs) of the parameters, based on

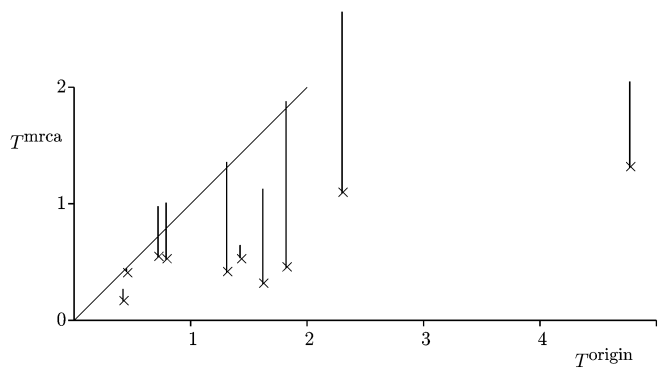

FIGURE 2. Scatter diagram of 10 realizations of the standardized joint distribution $(T^{\text{origin}}, T^{\text{mrca}}, Z)$. $T^{\text{origin}}$ is the time of origin, $T^{\text{mrca}}$ is the time of the most recent common ancestor, and $Z$ is the number of species at the time of most recent common ancestor. Points $\times$ give the $(T^{\text{origin}}, T^{\text{mrca}})$ values, and the length of the line segment is the $Z$ value. Note the extreme variability of $Z$; the smallest value was 0.03 and the largest was 1.56. Note that points are necessarily below the diagonal line $T^{\text{mrca}} = T^{\text{origin}}$.

TABLE 1. MLEs of linear birth–death parameters

| Realization | 1 | 2 | 3 | 4 | 5 | 6 | 7 | 8 | 9 | 10 |
|---|---|---|---|---|---|---|---|---|---|---|
| MLE of $\lambda$ | 0.4 | 0.6 | 1.3 | 1.5 | 0.6 | 1.3 | 0.3 | 0.3 | 0.9 | 1.2 |
| MLE of $\mu$ | 0.1 | 0.4 | 0.2 | 1.0 | 0.3 | 0.3 | 0.2 | 0.1 | 0.4 | 0.2 |
| MLE of $t_*$ | 9.1 | 6.5 | 1.7 | 3.1 | 4.7 | 2.5 | 11.4 | 13.8 | 4.2 | 1.5 |
| $T^{\mathrm{mrca}}$ | 8.4 | 6.0 | 1.5 | 2.8 | 4.3 | 2.4 | 11.3 | 13.3 | 3.9 | 1.4 |
| $T^{\mathrm{origin}}$ | 37.1 | 13.3 | 1.7 | 8.2 | 8.9 | 11.9 | 36.2 | 154.7 | 21.8 | 1.7 |

Notes: MLEs of linear birth–death parameters based on each of 10 realizations from the conditioned critical branching process model with $n=8$. Estimated parameters: $\lambda$ is the per-species speciation rate, $\mu$ is the per-species extinction rate, $t_*$ is the time before present of clade origin, $T^{\mathrm{mrca}}$ is the time of the last common ancestor, and $T^{\mathrm{origin}}$ is the time of origin.

a lineage tree as data. (The only subtle issue is that one should compute the likelihood *without* conditioning on $n$. To see why, note that when $\mu/\lambda$ is large, the process is a priori unlikely to reach $n$ species; this is a real effect that would incorrectly be factored out by conditioning.)

We studied what happens if one applies this procedure—estimating parameters assuming the underlying model of species diversity is a linear birth-and-death process—to simulated data from our (conditioned critical branching process) model. Of course, in our model we really have $\lambda = \mu = 1$ in each realization and realization-dependent values of $T^{\mathrm{mrca}}$ and $T^{\mathrm{origin}}$. Table 1 shows the MLEs derived from each of 10 typical realizations of the model, with $n = 8$.

So in this "small $n$" setting, the estimated values of $\lambda$ and $\mu$ in the linear model are very misleading. Not only does a "pull of the recent" effect make the estimated $\lambda$ larger than the estimated $\mu$ but also the estimated values are varying widely between realizations. However, it is quite possible that (unlike other variability effects in our model) these particular effects might diminish for larger $n$.

*Discussion*

The results above are not directly comparable with previous results such as Nee, Holmes et al. (1994), Nee, May, and Harvey (1994), Nee (2001), Paradis (1997, 1998, 2004), Rabosky (2006), and Rabosky and Lovette (2008), whose conclusions range from optimistic

> It is possible to estimate the rate of diversification of clades from phylogenies with a temporal dimension (Nee 2001, 661)

to equivocal

> ... the speciation rate was correctly estimated in a wide range of situations ... However, this estimator was substantially biased when the simulated extinction rate was high. On the other hand the estimator of extinction rate was biased in a wide range of situations. (Paradis 2004, 19)

See also Barraclough and Nee (2001) for discussion of practical difficulties. But to us, the results in Table 1 cast serious doubt on the ability to reconstruct at any level of detail the history of a single *small* clade from the phylogeny of extant species. In contrast, a statistical study of phylogenetic trees (with relative timescale) from *many* clades might provide some insight into typical patterns of recent speciation and extinction rates.

EXTANT ANCESTRAL SPECIES

Polar bears are often said to be a recent (perhaps 200 Kyr) daughter species of brown bear (DeMaster and Stirling 1981). This is an unusual (among familiar animals) instance where an ancestral species is extant and suggests the following.

**Question 3.** Within well-studied extant clades, what proportion $\alpha$ of extant species have some extant ancestor?

Anecdotally, biologists regard $\alpha$ as small, though we have been unable to find useful data, perhaps in part because cladistics dogma discourages asking this question. Our species model predicts (Aldous and Popovic 2005, Corollary 11) that for about 63% of extant species, some ancestral species should be itself extant (note that this is different from saying it has an extant descendant). Although this numerical value depends on arbitrary details of the particular model, an informal mathematical argument suggests that for any model incorporating extinctions and speciations that are not "tightly coupled" (in the sense that a speciation would usually be followed quickly by extinction of either daughter of parent species), the model will predict that some constant percentage (not close to 0%) of extant species will have extant ancestors. The argument is that a typical extant species has random age $T$; the chance $p$ that the direct ancestor species is extant would be (assuming independence—see below) the expectation of $g(T)$, where $g(t)$ is the chance that a species lifetime is at least $t$, and this quantity does not tend to 0 as $n \to \infty$. This calculation assumes independence of future species lifetimes and current speciation rate, which is only true in the simplest models, but to make $p \approx 0$, one would need almost complete dependence.

This contrast between models and data is striking to us, though biologists with whom we have discussed this issue tend to dismiss it as simply reflecting the practical difficulty in distinguishing similar species. See also the related recent work of Funk and Omland (2003) on species-level paraphyly. But our point is that all existing statistical study of questions involving past speciation or extinction rates implicitly depends on models of the type that predict nonsmall $\alpha$, and so all such work would appear less convincing if data really show $\alpha$ to be small.

## MODELING HIGHER ORDER TAXA

Our final two questions concern higher order taxa and involve a probability model for phylogenetic trees on, and diversity of, higher order taxa. In this section, we describe the model from Aldous et al. (2008). For concreteness of language, we write *genera*, but we mean any reasonable way of partitioning species into groups.

### Three Classification Schemes

Imagine a systematist is given a correct phylogeny on species and chooses a number of *apomorphies* (a trait that characterizes an ancestral species and its descendants; we call the ancestral species a *new type*) judged significant. Figure 3 illustrates three ways in which the same underlying data (a phylogeny on 22 species, of which 2 are new type) can be used to define genera and hence a tree on genera. We call the three schemes *fine*, *medium*, and *coarse*. Very roughly, one can interpret the fine scheme as attempting to conform with cladistic principles by making taxa be monophyletic, whereas the coarse scheme is what a traditional taxonomist given only fossil data showing these two apomorphies (and not knowing the true tree on species) might devise.

Here are some logical properties one might like such a classification scheme to possess.

**Property 1.** A genus cannot contain both a species $a$, which is a descendant of some "new type" species $s$, and also a species $b$, which is not a descendant of $s$.
Here "descendant" includes $s$ itself, so in particular a "new type" species and its parent must be in different genera.

Next note that if we required every genus to be a clade (monophyletic), then we could never have more than one genus because otherwise some parent–daughter pair $\{a,b\}$ would be in different genera and then the genus containing $a$ is not a clade. We will consider a weaker property. Any two distinct species $a,b$ have a *most recent common ancestor* MRCA$(a,b)$, which is some species (maybe $a$ or $b$). Given three distinct species $\{a,b,c\}$, say $(a,b)$ *are more closely related than* $(a,c)$ if MRCA$(a,b)$ is a descendant of MRCA$(a,c)$. Here again we allow MRCA$(a,b)$ = MRCA$(a,c)$.

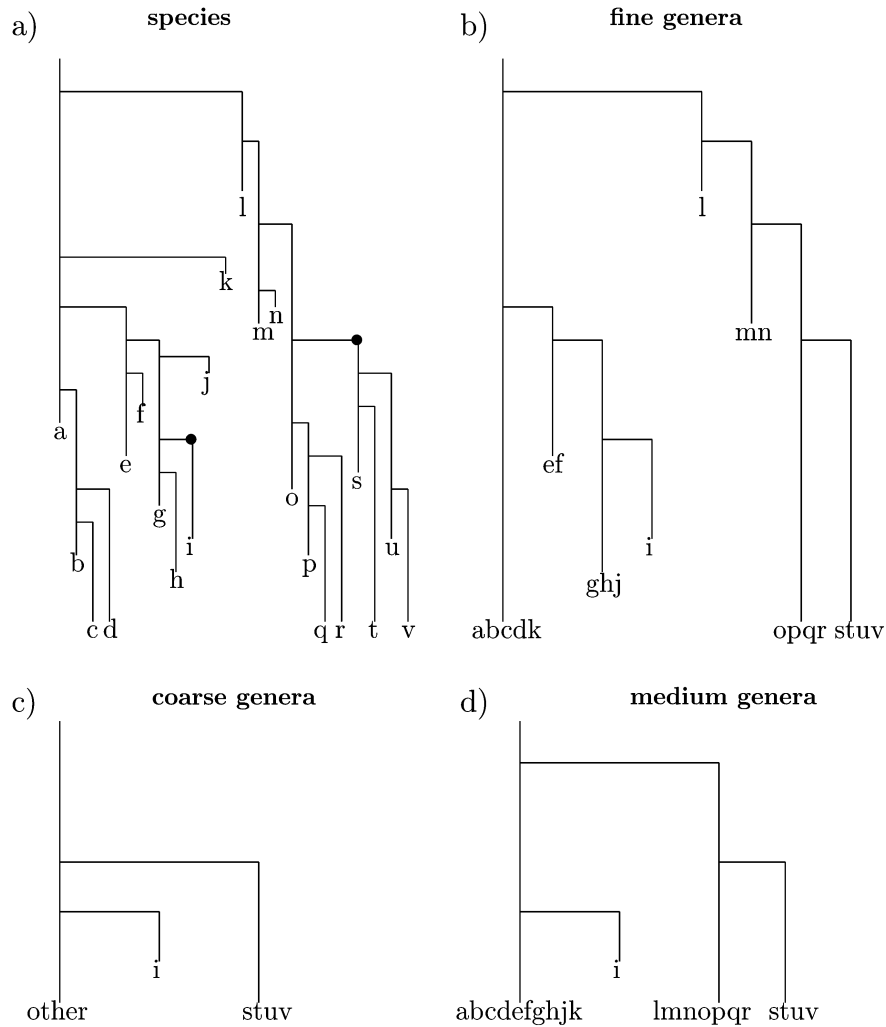

FIGURE 3. Illustration (from Aldous et al. 2008) of our schemes for defining genera in terms of new types. Above left is a complete clade (a) of 6 extant and 16 extinct species ($abcd\cdots uv$), with 2 species $\{i,s\}$ designated as new types and marked ●. In the fine scheme (b), this induces 8 genera (3 extant), whose tree is shown above right. The other schemes are shown below, with compressed timescale, giving in the coarse scheme (c) 3 genera and in the medium scheme (d) 4 genera.

**Property 2.** Given three distinct species $\{a,b,c\}$, with $a$ and $b$ in the same genus and $c$ in a different genus, then $(a,b)$ are more closely related than $(a,c)$.

As another kind of desirable property, one would like to be able to draw a tree or cladogram on *genera* in some unique way, and the next property (for a classification scheme) provides one formalization of this idea.

**Property 3.** Choosing one representative species from each genus and drawing the cladogram on these species give a cladogram that does not depend on the choice of representative species.

Note that the trivial scheme of assigning each species to a different genus possesses all these properties, so the issue is to find the *coarsest* schemes with such properties. It turns out that we can define three genera classification schemes (*coarse, medium, fine*) as the coarsest schemes with the following properties.

| Coarse | Property 1 |
| Medium | Properties 1 and 3 |
| Fine | Properties 1 and 2 |

Theorem 1 of Aldous et al. (2008) gives simple algorithmic descriptions of the constructions of the genera in each scheme. The coarse scheme is always at least as coarse as the other schemes. Proposition 2 of Aldous et al. (2008) shows that (except for one atypical possibility) the medium scheme is at least as coarse as the fine scheme. Because these three schemes seem mathematically and biologically reasonable, we did not investigate other possible schemes. We do not know any extensive previous work along these lines, though Scotland and Sanderson (2004) study one aspect of a model similar to our coarse scheme.

*Probability Models and Higher Taxa*

The classification schemes above do not involve probability but are one ingredient in the probability model we now describe. Start with the (conditioned critical branching process) species model. Declare each new species to be "new type" with probability θ (the parameter θ will indirectly determine the mean number of species per genus). Then choose one of the three classification schemes above.

To compare real data with the predictions of the model, one would do some appropriate conditioning (recall that in the species-level model we conditioned on the number $n$ of extant species). The model results we present in this paper do not depend on the details of such conditioning. As explained in Aldous et al. (2008) the model can be applied to extinct clades, but here we give results only for the case of extant clades.

PHYLOGENETIC TREE SHAPE AND
HIERARCHICAL LEVEL

It is a long-standing puzzle (Mooers and Heard 1997), with recent confirmation via exhaustive analysis of TreeBASE (Blum and François 2006), that real phylogenetic trees seem more "imbalanced" than predicted by a natural *Markov model*, though more balanced than a (less natural) *uniform* or *PDA* (*proportional to distinguishable arrangements*) model. Imbalance in a particular clade is often taken as evidence of some particular biological effect, as illustrated by the following quote:

> We combine statistical and phylogenetic approaches to test the hypothesis that adaptive radiation and key innovation have contributed to the diversity of the order Araneae. The number of unbalanced araneid clades (those whose species numbers differ by 90% or more) exceeds the number predicted by a null Markovian model. The current phylogeny of spider families contains 74 bifurcating nodes, of which 31 are unbalanced. As this is significantly more than the 14.8 expected unbalanced nodes, some of the diversity within the Araneae can be attributed to some deterministic cause (e.g., adaptive radiation). (Bond and Opell 1998, p. 403)

A statistical analysis of 61 phylogenies concluded

> Nodes with a given number of higher taxa descended from them were on average more unbalanced than were nodes with the same number of species as descendants. (Purvis and Agapow 2002, p. 844)

And in the different context of "deep phylogeny" relating extinct and extant groups, one frequently sees "comb" cladograms—see, for example, the phylogeny of terrestrial vertebrates by Laurin (2007)—in which single leaves split off one after another.

One can think of three interpretations of such cladograms. One possibility is that they are just wrong, that is, do not show the correct relationship between monophyletic terminal clades. An opposite possibility is that they are correct and show some significant biological effect, such as successive replacement or marginalization of "more primitive" groups by less primitive groups. An intermediate possibility is that such cladograms are simultaneously correct and artifactual, in the sense that a species-level tree would be more balanced, but the procedure of drawing cladograms whose terminal taxa are clades tends intrinsically to produce more unbalanced trees.

**Question 4.** When one moves from trees on species to trees on sets of species (whether traditional higher order taxa or clades within PhyloCode), does one expect trees to become more unbalanced as a purely logical consequence of tree structure, without signifying any real biological phenomenon?

### Measuring Tree Balance

To address Question 4 one needs to decide how to measure tree balance. There is a sizable literature (Purvis and Agapow 2002) on summary statistics $T_n$ measuring "balance" of an $n$-leaf tree and their distribution under the usual Markov model as a null hypothesis. Although this is the natural way to study individual trees, it is not satisfactory for studying the overall statistical properties of a collection of trees of different sizes because to compare different $n$ one needs to standardize and the standardization requires some null model, which begs the question.

A different method, used in Aldous (2001) on a few large trees, seems a less arbitrary and more powerful way of analyzing individual large trees or collections of small trees. Each branch point of a binary tree splits a clade of size $m$ (say) into subclades of sizes $a$ and $m - a$, where we take $a \leq m/2$ as the size of the smaller daughter clade. Given a collection of trees, take all the splits in all the trees and then calculate the function

$a(m)$ = average size of smaller daughter clade in split of size-$m$ clade.

Here we average over all size-$m$ clades within all trees in the collection—we are not comparing properties of different sized trees. This function provides a measure of "balance" in a collection of trees that has three advantages over using summary statistics (uses more within-tree structure; avoids arbitrary choice of summary statistic; avoids issues of normalization required to compare different size trees). A companion function, useful in the context of studying occurrence of comb-like cladograms, is

$p(m)$ = proportion of splits of size-$m$ clades where smaller daughter clade size $= 1$.

### Predictions of Our Model

Table 2 shows the measures $a(m)$ and $p(m)$ of tree balance within our model.

Recall that we are talking about trees whose terminal taxa are the genera. Within such trees, we look at clades consisting of $m = 5, 10, 15$ genera, and for the root split of such clades, we record the mean size $a(m)$ of the smaller subclade and the proportion $p(m)$ of smaller subclades with size 1. This is repeated for different values of the parameter $\theta$ chosen to make the mean number of species per genus to be $\mu = 1, 5, 10, 20$. The first row ($\mu = 1$) is just the species-level model, where our model coincides with the usual Markov model and predicts uniformly distributed splits. Increasing imbalance as $m$ increases

TABLE 2. Shape of trees on extant genera

| | Parent clade size $m$ | | | | | | |
|---|---|---|---|---|---|---|---|
| | 5 | | 10 | | 15 | | |
| $\mu =$ mean number species per genus | $a(5)$ | $p(5)$ | $a(10)$ | $p(10)$ | $a(15)$ | $p(15)$ | $\theta$ |
| Coarse | | | | | | | |
| 1 | 1.50 | 0.50 | 2.77 | 0.22 | 4.02 | 0.15 | 1 |
| 5 | 1.42 | 0.58 | 2.51 | 0.30 | 3.67 | 0.19 | 0.198 |
| 10 | 1.35 | 0.65 | 2.36 | 0.35 | 3.59 | 0.19 | 0.100 |
| 15 | 1.31 | 0.69 | 2.28 | 0.38 | 3.5 | 0.19 | 0.066 |
| 20 | 1.28 | 0.72 | 2.34 | 0.35 | 3.64 | 0.20 | 0.050 |
| Medium | | | | | | | |
| 1 | 1.50 | 0.50 | 2.77 | 0.22 | 4.02 | 0.15 | 1 |
| 5 | 1.54 | 0.46 | 2.72 | 0.24 | 3.91 | 0.16 | 0.136 |
| 10 | 1.52 | 0.48 | 2.69 | 0.24 | 3.87 | 0.17 | 0.061 |
| 15 | 1.51 | 0.49 | 2.66 | 0.25 | 3.91 | 0.17 | 0.040 |
| 20 | 1.50 | 0.50 | 2.67 | 0.26 | 3.83 | 0.19 | 0.029 |
| Fine | | | | | | | |
| 1 | 1.50 | 0.50 | 2.77 | 0.22 | 4.02 | 0.15 | 1 |
| 5 | 1.26 | 0.74 | 2.55 | 0.34 | 3.78 | 0.21 | 0.042 |
| 10 | 1.17 | 0.83 | 2.40 | 0.41 | 3.71 | 0.25 | 0.011 |
| 15 | 1.13 | 0.87 | 2.18 | 0.49 | 3.47 | 0.31 | 0.005 |
| 20 | 1.12 | 0.88 | 2.04 | 0.54 | 3.68 | 0.38 | 0.003 |

Notes: For the given size (number of genera) $m$ in a parent clade, the table shows the probability $p(m)$ that smaller daughter clade size equals 1 and the mean size $a(m)$ of smaller daughter clade. Results from Monte Carlo simulations of model with 200 extant species. For $\mu = 1$, the true values of $a(15)$ and $p(15)$ are 4 and $1/7 = 0.143$; the values shown indicate the (small) errors from simulation.

would be indicated by $p(m)$ increasing and by $a(m)$ decreasing.

So our model predicts the following:

1. Imbalance increases with size of genus, that is, as we go up the taxonomic hierarchy;
2. the increase in imbalance is most prominent for the fine scheme and least prominent for the medium scheme.

This analysis is consistent with the possibility that observed imbalance in trees on higher level taxa may be in part an artifact of classification. On the other hand, two studies (Heard 1992; Mooers 1995), based on summary statistics of published small trees, discussed in Mooers and Heard (1997), conclude that there is no such hierarchical trend in imbalance. It would be interesting to repeat such data analysis on larger trees.

FLUCTUATIONS AT DIFFERENT HIERARCHICAL LEVELS

The compendia by Sepkoski (1992) are justly celebrated for providing raw data for the statistical study of long-term evolutionary history (see also Benton 1993). Because of the difficulty of resolving fossils to the species level, such data are typically presented as time series for numbers of genera and families, raising the issue of how reliable is it as a proxy for time series for numbers of species. Paleontologists tend to regard it as reliable, as indicated in the first sentence below and confirmed from data such as Lane and Benton (2003).

> The complex trajectory of taxonomic diversity through [600 myr] has proved robust to continued sampling and, as shown by simulations, to very different phylogenetic approaches to grouping species into higher taxa. But diversity time series become increasingly jagged and disparate at lower taxonomic levels and on regional scales, both because sampling is less complete and because *lower-diversity lineages really are almost inevitably more volatile*. (Jablonski 1999, p. 2114; our emphasis added)

We focus on the emphasized assertion: is this *inevitability* a biological or a mathematical effect?

**Question 5.** How do we expect that volatility for counts of higher order taxa should compare with volatility for number of species, as a purely mathematical effect?

Suppose we have data, for a sequence of times $\tau(0), \tau(1), \ldots, \tau(k)$, on number of families at each time and numbers of genera in each family at each time. There is a technical issue of how to measure *volatility* as a "standardized fluctuation rate" to permit fair comparisons. After deciding that, within our model we can see how volatilities do vary between levels, as a purely mathematical effect.

*Standardized Fluctuation Rates*

We now take the basic species-level model (conditioned critical branching process) but (now thinking of extinct clades) do not condition on number of extant species. Write $N(t)$ for number of species at time $t$ and $G(t)$ for number of genera at time $t$, using one of our schemes for defining genera, and recall that the time unit is mean species lifetime. A basic mathematical property of our model is that the stochastic fluctuations, measured by variance of changes in the time series, have a simple form over time intervals whose duration $t$ is of the same order as mean species lifetime. Given $N(t_0) = n(0)$, we have

$$\text{var}(N(t_0 + t) - n(0)) \approx 2n(0)t.$$

Intuitively, this holds because

$$\text{given } N(t) = n(t), \quad \frac{d}{dt}\text{var}N(t) = 2n(t)$$

and because $N(t) = n(0) \pm O(\sqrt{n(0)})$ over times of order mean species lifetime. Translating into real time units $\tau$ (measured in myr, say) and writing $\mu_s$ for mean species lifetime,

$$\text{var}(N(\tau_0 + \tau) - n(0)) \approx 2n(0)\tau/\mu_s.$$

In other words, the ratio

$$\frac{\mu_s \text{var}(N(\tau_0 + \tau) - n(0))}{2n(0)\tau} \quad (1)$$

is approximately 1 regardless of the value of $n(0)$ (assumed not too small) or the value of $\tau$ (assumed of order $\mu_s$).

This analysis suggests a way to define standardized fluctuation rates for time series of genera. Write $\mu_g$ for mean genus lifetime. If we were to model genera directly as behaving statistically like species, then the standardized fluctuation rate for genera, defined below by copying Equation 1, would equal 1. Given $G(\tau_0) = g(0)$, define the analog of Equation 1 as

standardized fluctuation rate for genera

$$= \frac{\mu_g \text{var}(G(\tau_0 + \tau) - g(0))}{2g(0)\tau} \quad (2)$$

for $\tau$ of order $\mu_g$. So fluctuation rate for genera that turns out to be different from 1 is indicating a mathematical effect of working with genera composed of species, rather than genera envisaged as autonomous entities. Table 3 gives numerical values predicted by our model.

Returning to the original question, the point is that our model treats different higher taxonomical levels (genera and families, for concreteness) in the same way, simply using two different parameter values to fit mean number of species per genus and mean number of species per family. So from the kind of results shown in Table 3, one can derive relative volatilities. For instance, if the data showed an average of 4 genera per family, and if we were willing to guess an average of 5 species per genus, then Table 3 shows that (within our model) the fluctuation rate for families relative to the fluctuation rate for genera would be

$0.68/0.84 = 0.81$ [coarse]; $\quad 0.89/1.08 = 0.82$ [medium]; $0.50/0.95 = 0.53$ [fine].

Of course, it would be interesting to try to estimate such ratios in paleontological data.

*Previous Work*

Previous work such as Sepkoski and Kendrick (1993) and Robeck et al. (2000) had a rather different focus, motivated by the incompleteness of the fossil record. What classification scheme, applied to data from randomly sampled species, enables one to most accurately estimate the underlying trends in species diversity?

TABLE 3. Standardized fluctuation rates for genera

| Mean number species per genus | 1 | 5 | 10 | 15 | 20 |
|---|---|---|---|---|---|
| Coarse | | | | | |
| Standardized fluctuation rate | 1.00 | 0.84 | 0.72 | 0.70 | 0.68 |
| Mean genus lifetime | 1.00 | 1.89 | 2.35 | 2.66 | 2.85 |
| Medium | | | | | |
| Standardized fluctuation rate | 1.00 | 1.08 | 1.03 | 0.94 | 0.89 |
| Mean genus lifetime | 1.00 | 2.44 | 3.52 | 4.24 | 4.90 |
| Fine | | | | | |
| Standardized fluctuation rate | 1.00 | 0.95 | 0.71 | 0.58 | 0.50 |
| Mean genus lifetime | 1.00 | 2.41 | 3.26 | 3.82 | 4.21 |

Notes: Results from Monte Carlo simulations of model started with 200 species. Ratio (2) estimated with $\tau$ = mean genus lifetime.

That previous work studied a variety of schemes that (roughly speaking) started with what we have called the coarse scheme but then applied various ad hoc methods of dealing with resulting paraphyletic groups; they also assumed logistic or exponential clade expansion and occasional mass extinctions. It would be interesting to study performance of our classification schemes under the latter assumptions.

Of course, it is easy to criticize any scheme as "not what taxonomists actually do", but it is not clear that our schemes are less realistic than those used in previous work.

GENERAL DISCUSSION

We should reemphasize that our purpose is not to propose models realistic enough to use directly in the analysis of real data, but rather to set out a general modeling framework. Mathematical models of real-world phenomena can be placed on a spectrum, with "crude toy models" or "thought experiments" at one end of the spectrum and models that claim to give numerically accurate predictions (testable by experiment or statistical analysis of data) at the other end. Our models should be placed firmly at the "thought experiment" end, and indeed the reader will have noticed that some of the questions we pose seem unanswerable with known data. The novel feature of the models is that they treat different taxonomic levels, tree shape and time series, extinct and extant clades, all in a logically consistent way. This contrasts with literature modeling time series of genera separately as a random walk (Gould et al. 1977) or a birth-and-death process (Stoyan et al. 1983).

We used a "neutral" species-level model for two reasons. One is mathematical simplicity. Our model has one parameter (mean species lifetime). The linear birth-and-death model has two parameters, but diversity cannot grow exponentially forever, so to force logistic-type behavior one needs a third parameter, representing, for example, ecological constraints on diversity. Such models have been studied in recent literature on quantitative aspects of adaptive radiation, such as Rabosky and Lovette (2008), Phillimore and Price (2008), and Rabosky (2009), which implicitly focus on estimation of speciation and extinction rates within individual large clades. We do not have anything to contribute to such studies at a technical level but offer two observations. First, technical studies assume that some particular probability model is true. The exercise (following Question 2) of repeating an analysis on hypothetical data from a different model provides a useful reminder that conclusions may be surprisingly sensitive to assumptions. Second, existing studies such as those cited above and Mooers and Heard (1997), Purvis and Agapow (2002) and Webster et al. (2003) have been done in the context of testing some specific statistical hypothesis. But a useful parallel project would be to work toward some overall statistical description of histories of typical clades.

Of course, our methodology for extension to higher order taxa could be applied to the potentially more realistic multiparameter species models. Branching process theory suggests the following behavior in the supercritical ($\lambda > \mu$) setting. There will be two qualitatively different cases. If the growth rate $\lambda/\mu$ of number of species is sufficiently large compared with the parameter $\theta$ controlling rate of formation of new genera, then there will be large genera—that is, the largest genus will contain a nonzero proportion of all $n$ extant species, for large $n$. The other case, when $\lambda/\mu$ is closer to 1, will be more similar to the neutral case in this paper, in that the genera size distribution will have a limit for large $n$.

FUNDING

This work was supported by the National Science Foundation (grant number DMS0704159).

ACKNOWLEDGMENTS

We have had brief correspondence with many biologists concerning this ongoing project, and we thank those who have made helpful comments: Steven Heard, Susan Holmes, Graeme Lloyd, Jeffrey McKee, Arne Mooers, Mark Pagel, Mike Steel, and two anonymous reviewers.